# Faint-state transitions in the SW Sextantis nova-like variable, HS 0455+8315

Jeremy Shears, Boris Gänsicke, Pablo Rodríguez-Gil, David Boyd, Graham Darlington & Ian Miller

## Abstract

We present the fourteen year-long light curve of the SW Sextantis nova-like variable, HS 0455+8315, from 2000 November to 2015 February which reveals two deep faint states at magnitude 19 – 20, each of which lasted about 500 and 540 days. Outside these faint states, the star spent most of the time in a normal state at a magnitude of about 15.3. The second faint state was the better observed of the two and was found to have a linear decline of 0.009 mag/day, which was soon followed by a more rapid brightening at -0.020 mag/day. Time series photometry during both the normal state and near minimum light at magnitude ~18 showed that the eclipses had very similar profiles and that outside the eclipse there were irregular modulations typical of the flickering inherent to accreting CVs. Our photometry leading up to the minimum shows that accretion was still ongoing during this time.

## Introduction

The SW Sex stars are a sub-class of nova-like cataclysmic variables (CVs) originally proposed by Thorstensen *et al.* in 1991 (1) and later reviewed by Rodríguez-Gil *et al*. in 2007 (2). It initially comprised only eclipsing nova-like stars with typical orbital periods of 3 to 4 h and which exhibit single-peaked emission lines, strong He II emission and transient absorption features at orbital phase 0.5 irrespective of the inclination (2) (3); the SW Sex class was later extended to lower orbital inclination non-eclipsing systems which display the same spectroscopic characteristics. SW Sex stars have high luminosities and hot white dwarfs, implying extremely high accretion rates (4). Recently it has been suggested that SW Sex stars represent the dominant CV population in the 3 to 4 h orbital period range, which, if true, implies that the SW Sex phenomenon is likely to be an evolutionary stage in the life of a CV (2) (5). This makes the study of SW Sex stars particularly important to our understanding of CV evolution.

Some SW Sex stars show occasional faint states when mass transfer is reduced or even completely stopped resulting in a drop in brightness of 3 – 5 magnitudes. The stars can stay at these low levels for weeks or months before rising again to their normal state. The reasons for these faint states are not completely understood. It is therefore important that low states are observed, to determine what is happening in the system on the approach to and during the faint state. The compendium of known or possible SW Sex stars continues to grow and the current “Big List” is maintained online by Don Hoard (6). Approximately one-third of stars on the list have been observed to exhibit low states.

HS 0455+8315 was identified as an SW Sex star by Gänsicke et al. (7) during follow up observations of optically selected CV candidates from the Hamburg Quasar Survey (HQS) (8). They found that out of eclipse the system had R ≈ 14.7 with moderate flickering activity (up to 0.3 magnitude full amplitude) and deep eclipses (ΔR ≈ 1.5). Subsequent analysis of the times of mid-eclipse revealed $P_{orb}$ = 3.569 h (2). A refined eclipse ephemeris was published by Boyd (9), who also showed the period was constant to 1 part in $2\times10^6$ over an interval of ten years.

HS 0455+8315 lies in Cepheus at RA 05 06 48.27 Dec +83 19 23.3 (J2000.0). It is listed in the USNAO-A2.0 as object 1725-00218395 (magnitude 14.8 from a blue photographic plate and 14.6 from a red plate) and in the 2MASS catalogue as 2MASS J05064821+8319235.

In this paper we present the fourteen-year light curve of HS 0455+8315 which reveals two faint states.

**Observations**

To produce a long-term light curve of HS 0455+8315, we used data from the BAA VSS database, the AAVSO International Database, the Center for Backyard Astrophysics database and the Spanish M1 group. The vast majority of these data are unfiltered CCD measurements with V-band comparison stars, although a few are V-band CCD measurements. The observers whose data were used are listed in Table 1, in addition to the authors' data.

We obtained R-band CCD photometry using the IAC-80 0.82 m telescope located at the Observatory del Teide on Tenerife to investigate the eclipses. A Thomson 1 k x 1k pixel CCD camera was used; the comparison stars are the ones shown in reference (2) whose USNO-A2.0 R-band magnitudes are given in Table 2 of that paper.

We also obtained unfiltered CCD photometry before and during the 2013/14 low state using a distributed network of smaller telescopes. The instrumentation used is shown in Table 2 and the observation log in Table 3.

**Results**

*Long-term light curve*

We removed the eclipses from the light curve by subtracting data within 0.08 orbital periods of the times of minimum predicted by Boyd's ephemeris (9) and took the average brightness for each night. The resulting fourteen-year light curve of HS 0455+8315, from 2000 November to 2015 February is shown in Figure 1. We note that the data prior to early 2007 are sparse, but during the remaining part of the light curve the cadence was higher, with a mean interval between observations of 7 days and a median of 4 days. The increase in observations was stimulated by the publication by Gänsicke during 2006 of a web page encouraging amateur observers to monitor a range of CVs from the HQS (10).

Two deep fadings to a faint state at magnitude 19 - 20 were observed. The first event was less well defined by the available data, but appears to have lasted about 500 days from 2005 July to 2006 November and with a minimum in the middle of 2006 January. The second fading event lasted about 540 days, from 2013 February to 2014 August and with a minimum in early 2014. We note an apparent rise above the mean level following the second fading. During the interval between the two fadings, the star was in its normal state during which the light curve was remarkably flat with an average magnitude of 15.3 and a standard deviation of 0.2 magnitudes. Some of this spread might reflect a real variation in the star, but we cannot be sure since different photometric regimes were used, e.g. different detectors, filters and comparison stars.

An expanded plot of the second fading event appears in the top panel of Figure 2 which shows it to be asymmetric. The transition from normal to faint and from faint to normal are indicated by the blue lines in Figure 2. The brightening transition, -0.020 mag/d, was about twice as fast as the fading transition, 0.009 mag/d. These values correspond to *e*-folding times, $\tau$, of $\tau_{rise}$ = 54 and $\tau_{fall}$ =121 days, where $\tau = ((\log_{10} e)/0.4)/(\Delta mag/\Delta t)$.

It is difficult to be certain how long the star was at minimum, since there is much scatter in the data resulting from the photometric errors which are typically 0.14 mag at this time and are due to the low signal-to-noise ratio of the faint object. We conclude that it was no more than 50 days, but it is also possible that the brightening began as soon as the fade ended. Another problem is that, due to the incomplete time sampling, we cannot say for sure that the system's brightness hit the low state before the start of the plateau we see at ~18.5, and then experienced a series of re-brightenings ending again at the low state level.

*Eclipses during the normal and faint states*

Figures 3 and 4 show R-band eclipse profiles obtained with the IAC-80 telescope during the normal state in 2000 and near the bottom of the fading towards the faint state in 2013.

Analysis of the eclipses was complicated by out-of-eclipse variations in brightness. We therefore drew a baseline corresponding to the average light curve before and after the eclipse and then measured the depth and the full width at half minimum (FWHM) duration of each eclipse (Table 4). The three normal state eclipses had an average depth of 1.6 mag and a mean duration of 12.1 min, whereas the values for the faint state eclipse were 1.6 mag and 13.4 mins respectively. We conclude from this, and from a visual inspection of the eclipse profiles, which are symmetrical, that there is no significant difference between normal and faint state eclipses.

We attempted to investigate the eclipse depth and duration using the unfiltered photometry. This was problematic due to the low signal-to-noise ratio and time resolution obtained with the small telescopes available to us, which meant that in several cases the eclipse profiles were insufficiently constrained to measure their depth and duration. We should thus treat the measured eclipse parameters with caution. However, as with the R-band data, we could not establish a difference in the eclipses before the fade and during the approach to the faint state (Figure 2 and Table 4). Examples of some of the longer runs showing some eclipses are shown in Figure 5.

Outside the eclipses, irregular short-term modulations with peak-to-peak amplitude up to 0.3 magnitudes were observed in the IAC-80 R-band data (Figure 3 and 4). Even larger variations of up to 0.5 magnitudes are present in the unfiltered data taken with the smaller telescopes (Figure 5). We analysed the data from each individual run (as well as the combined data from the normal state and then from the faint state), having excluded the eclipses, using the Lomb-Scargle and ANOVA algorithms in the Peranso V2.50 software (11), but could not find a significant stable period in the power spectra. Visual inspection of the light curves shows that the humps occur in an irregular manner with an interval between them of 12 to 25 mins. This could be large amplitude flickering although we cannot be certain.

## Discussion

The results presented here clearly show two normal state/faint state transitions typical of nova-like CVs in the 3 to 4 h orbital period range where the amplitude was about 5 magnitudes. Honeycutt and Kafka (12) found that the transitions to and from faint states exhibit some systematic properties. Most of these transitions were adequately fit by a single straight line to the magnitude light curve and most showed faster rises than falls, such as we found for HS 0455+8315. They found that the average *e*-folding time was about 20 days, although there was a considerable variation with $\tau$ = 3 to 94 days. We note that the $\tau_{rise}$ we measured for HS 0455+8315 is within this range, whereas its $\tau_{fall}$ is somewhat longer.

As noted in the Introduction it appears that accretion may sometimes switch off completely during the faint state. Whilst our photometry was conducted just before the minimum of the faint state, rather than during the minimum itself, we conclude that at this time significant accretion was still occurring. The fact that the eclipse profile was similar to normal state eclipses, and that the eclipses did not become shorter during the 2013/4 fading, suggests that the accretion disc was of similar size in both states. If accretion had switched off we might have expected to see narrower eclipses than in the normal state, with a steeper component corresponding to white dwarf eclipse ingress and egress, and perhaps ellipsoidal modulation. Moreover, we note that the out-of-eclipse modulations, which are typical of accreting CVs, were present in both states. These are most likely associated with flickering due to the stochastic nature of the accretion process. Flickering has generally been observed during low states, although it does occasionally disappear, such as in the case of MV Lyr at extreme minimum (13).

Whilst we observed eclipses of different duration and depth, we cannot correlate these with the system state. One might expect the accretion disc to shrink during a low state, resulting in a shorter eclipse, but we have no evidence for this. A change in the duration of an eclipse can also be observed if the disc is eccentric. Indeed, such variations have been observed in the SW Sex system DW UMa which are attributed to an elliptical disc and where both eclipse duration and amplitude are correlated with the precession period of the disc (14).

It is noteworthy that much of the long-term photometry reported in this paper was obtained by amateur observers with small telescopes, which highlights the continuing value of amateur astronomers' observations even in the era of wide-field transient surveys. In any case, the high northerly declination of this object puts it out of the range of some surveys. It is thus important to continue monitoring HS 0455+8315 to determine how frequently it undergoes faint states and to what extent these faint states vary in duration and brightness. It is possible that further observations of eclipses during faint states might reveal times when accretion is at a low level, or ceases altogether, at which time eclipse mapping with professional-class instrumentation might reveal the white dwarf and spectroscopy might shed further light on the nature of the secondary star.

## Conclusions

The fourteen year light curve of HS 0455+8315 from 2000 November to 2015 February shows that the star spends most of the time at a normal state with a magnitude of about 15.3. However, two deep fadings to magnitude 19 - 20 were observed, one in 2005-6 and one in 2013-14, which lasted about 500 and 540 days respectively. The second event was the better observed and was found to have a linear decline of 0.009 mag/day which was soon followed by a more rapid brightening at -0.020 mag/day. This behaviour is typical of nova-like systems in the 3 to 4 hour orbital period range, such as been observed in many other SW Sex-type CVs.

Time series photometry during the normal state and near the minimum showed that the eclipses had very similar profiles and that there were irregular out-of-eclipse modulations with peak-to-peak amplitude up to 0.5 magnitudes. This behaviour is typical of the flickering inherent to accreting CVs. We conclude that accretion was still occurring at least during the approach to minimum.

**Acknowledgments**

The authors gratefully acknowledge the use of observations from the databases of the BAA VSS, the AAVSO, the Center for Backyard Astrophysics (data kindly supplied by Jonathan Kemp) and the Spanish M1 group. It is largely thanks to the latter group that the 2005/6 fade was observed. This article is based in part on observations made with the IAC80 telescope operated on the island of Tenerife by the Instituto de Astrofísica de Canarias in the Spanish Observatorio del Teide. The research made use of data from SIMBAD, operated through the Centre de Données Astronomiques (Strasbourg, France), and the NASA/Smithsonian Astrophysics Data System.

**Addresses**

Shears: “Pemberton”, School Lane, Bunbury, Tarporley, Cheshire, CW6 9NR, UK [bunburyobservatory@hotmail.com]

Gänsicke: Department of Physics, University of Warwick, Coventry, UK

Rodríguez-Gil: Instituto de Astrofísica de Canarias, Vía Láctea s/n, La Laguna, E-38205, Santa Cruz de Tenerife, Spain and Departamento de Astrofísica, Universidad de La Laguna, Avda. Astrofísico Francisco Sánchez, s/n, La Laguna, E-38206, Santa Cruz de Tenerife, Spain

Boyd: 5 Silver Lane, West Challow, Wantage, Oxon, OX12 9TX, UK

Darlington: Thakeham Observatory, West Sussex, UK

Miller: Furzehill House, Ilston, Swansea, SA2 7LE, UK

| **Name** | **Organisation** |
|---|---|
| Sensi & Jose Antonio | M1 |
| Josep M Bosch | M1 |
| Lew Cook | CBA |
| Michael Cook | AAVSO |
| Judá Curto | M1 |
| Adolfo Darriba Martinez | AAVSO/M1 |
| William Goff | AAVSO |
| Tom Krajci | AAVSO |
| Stuart Littlefair | AAVSO |
| Daniel José Mendicini | M1 |
| David Messier | CBA |
| Panos Niarchos | CBA |
| Miguel Rodríguez Marco | M1 |
| Diego Rodriguez Perez | AAVSO/M1 |
| Javier Ruiz | M1 |
| Richard Sabo | AAVSO |
| Donn Starkey | CBA |
| Observatorio de Cantabria | M1 |

**Table 1: Observers contributing data (in addition to the authors)**

AAVSO= American Association of Variable Star Observers, M1 = Spanish M1 group, CBA = Center for Backyard Astrophysics

| Observer | Telescope | CCD |
|---|---|---|
| Boyd | 0.35m SCT | Starlight Xpress SXV-H9 |
| Darlington | 0.35m SCT | SBIG ST9-XE |
| Miller | 0.35m SCT | Starlight Xpress SXVR-H16 |

**Table 2: Instrumentation used for time series photometry**

| Date (UT) | Start time (JD) | End time (JD) | Duration (h) | Observer |
|---|---|---|---|---|
| 2012 November 05 | 2456237.265 | 2456237.509 | 5.9 | Darlington |
| 2012 November 10 | 2456242.415 | 2456242.474 | 1.4 | Darlington |
| 2012 November 23 | 2456255.451 | 2456255.535 | 2.0 | Darlington |
| 2012 December 09 | 2456271.416 | 2456271.446 | 0.7 | Boyd |
| 2012 December 12 | 2456274.396 | 2456274.435 | 1.0 | Boyd |
| 2013 September 02 | 2456538.379 | 2456538.411 | 0.8 | Boyd |
| 2013 October 09 | 2456575.356 | 2456575.532 | 4.2 | Miller |
| 2013 October 10 | 2456576.328 | 2456576.615 | 6.9 | Miller |
| 2013 October 12 | 2456578.263 | 2456578.499 | 5.7 | Darlington |
| 2013 October 14 | 2456579.526 | 2456579.659 | 3.2 | Miller |
| 2013 November 30 | 2456627.262 | 2456627.411 | 3.6 | Miller |
| 2013 December 09 | 2456636.237 | 2456636.276 | 0.9 | Boyd |

**Table 3: Log of time series photometry leading up to and during the 2013/4 fade**

| Eclipse date (JD) | Eclipse date (UT) | System state | Filter | Depth (mag) | Duration FWHM (mins) |
|---|---|---|---|---|---|
| 2451859.2 | 2000 November 10 | Normal | C | 1.5 | 12.1 |
| 2451865.3 | 2000 November 16 | Normal | C | 1.6 | 10.7 |
| 2451884.2 | 2000 December 10 | Normal | C | 1.6 | 13.4 |
| 2456237.4 | 2012 November 05 | Normal | R | 2.2 | 9.4 |
| 2456242.4 | 2012 November 10 | Normal | R | 2.4 | 9.6 |
| 2456255.5 | 2012 November 24 | Normal | R | 1.5 | 8.3 |
| 2456271.4 | 2012 December 09 | Normal | R | 1.5 | 14.2 |
| 2456274.4 | 2012 December 12 | Normal | R | 1.6 | 13.5 |
| 2456538.4 | 2013 September 02 | Approaching faint | R | 2.2 | 13.3 |
| 2456580.6 | 2013 October 15 | Approaching faint | C | 1.6 | 13.4 |

**Table 4: Eclipse characteristics measured from IAC-80 R-band photometry and unfiltered photometry**

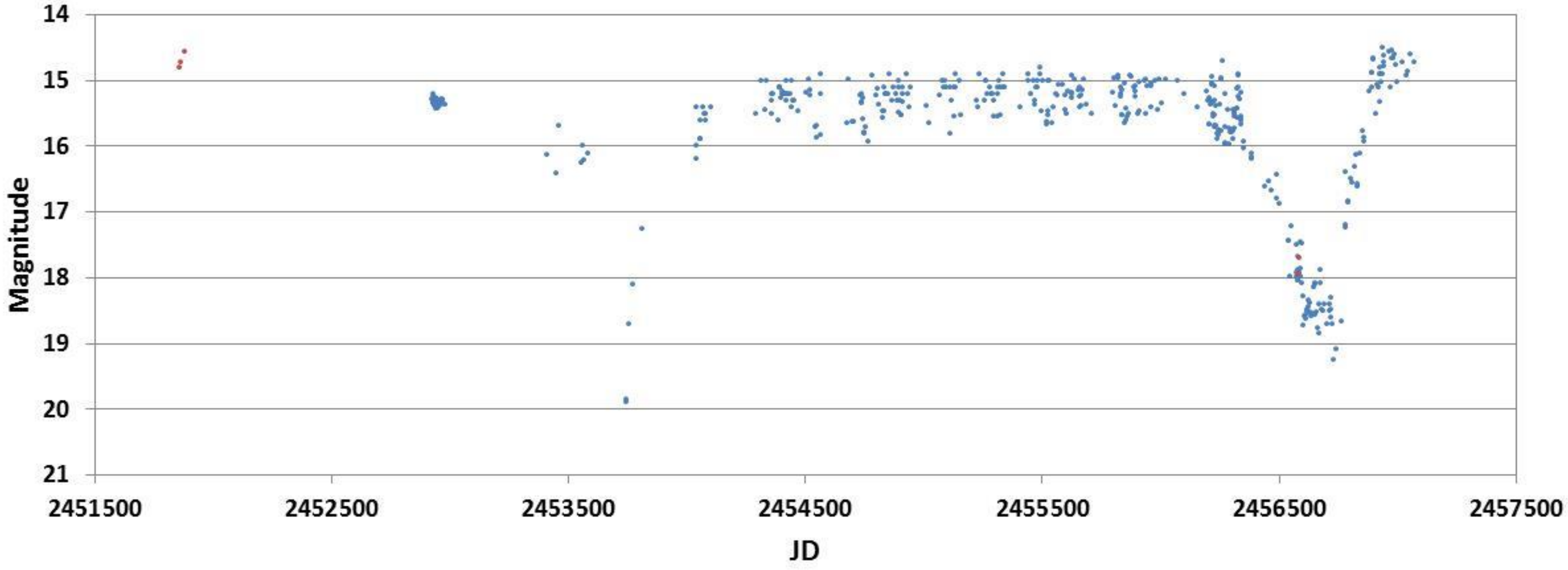

Figure 1: Light curve of HS 0455+8315, from 2000 November to 2015 February

Blue data points are C and V-band data; red data points are R-band

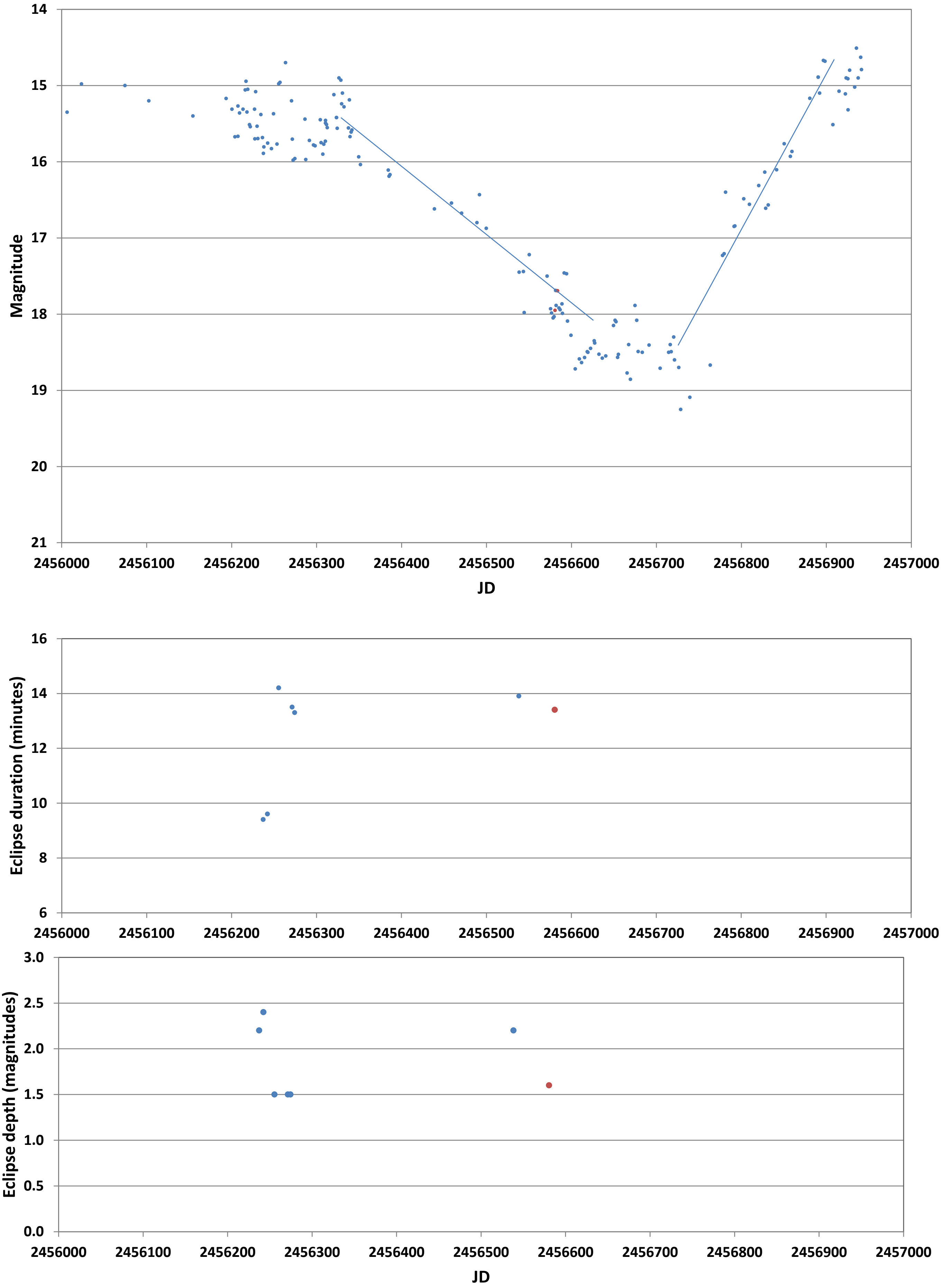

**Figure 2: Expanded view of the second fading event.** Top: out-of eclipse photometry (the blue lines indicate a linear fit to the transitions). Middle: eclipse duration. Bottom: Eclipse depth (FWHM). Blue data points are C and V-band data; red data points are R-band.

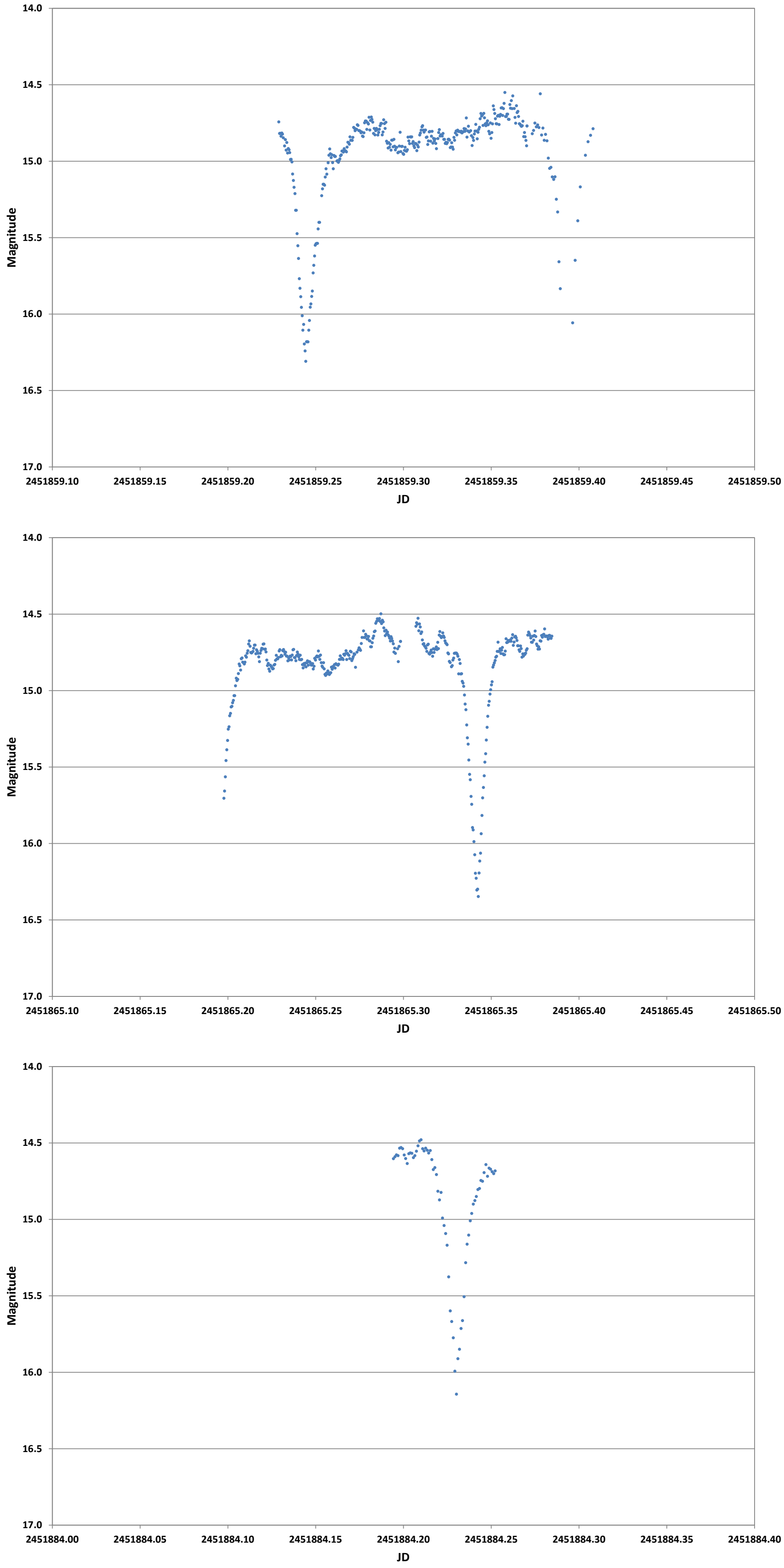

**Figure 3: Eclipses during normal state in 2000**

Top: November 10. Middle: November 16. Bottom: December 5

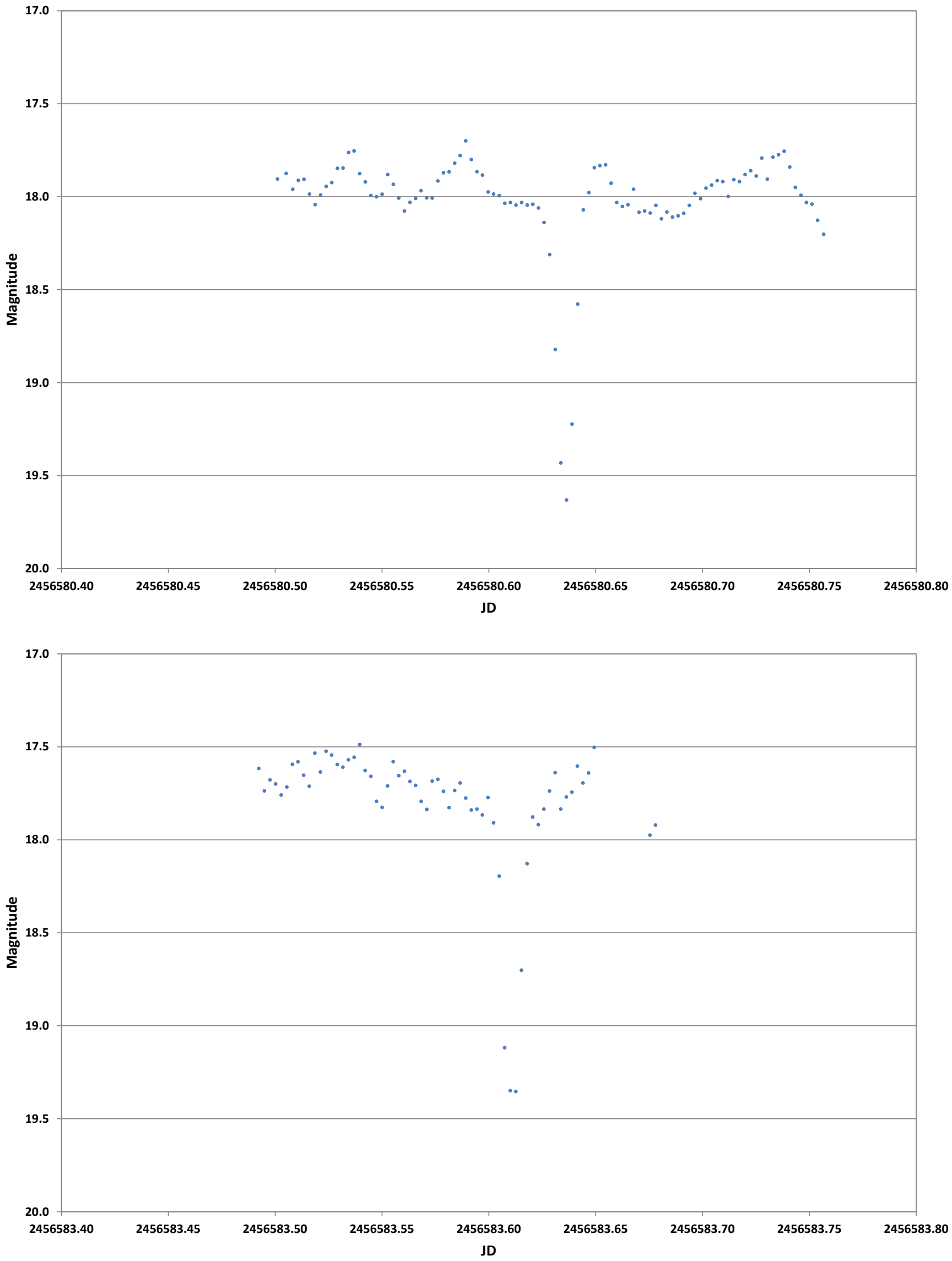

**Figure 4: Eclipses during faint state on 2013**

Top: October 14. Bottom: October 17

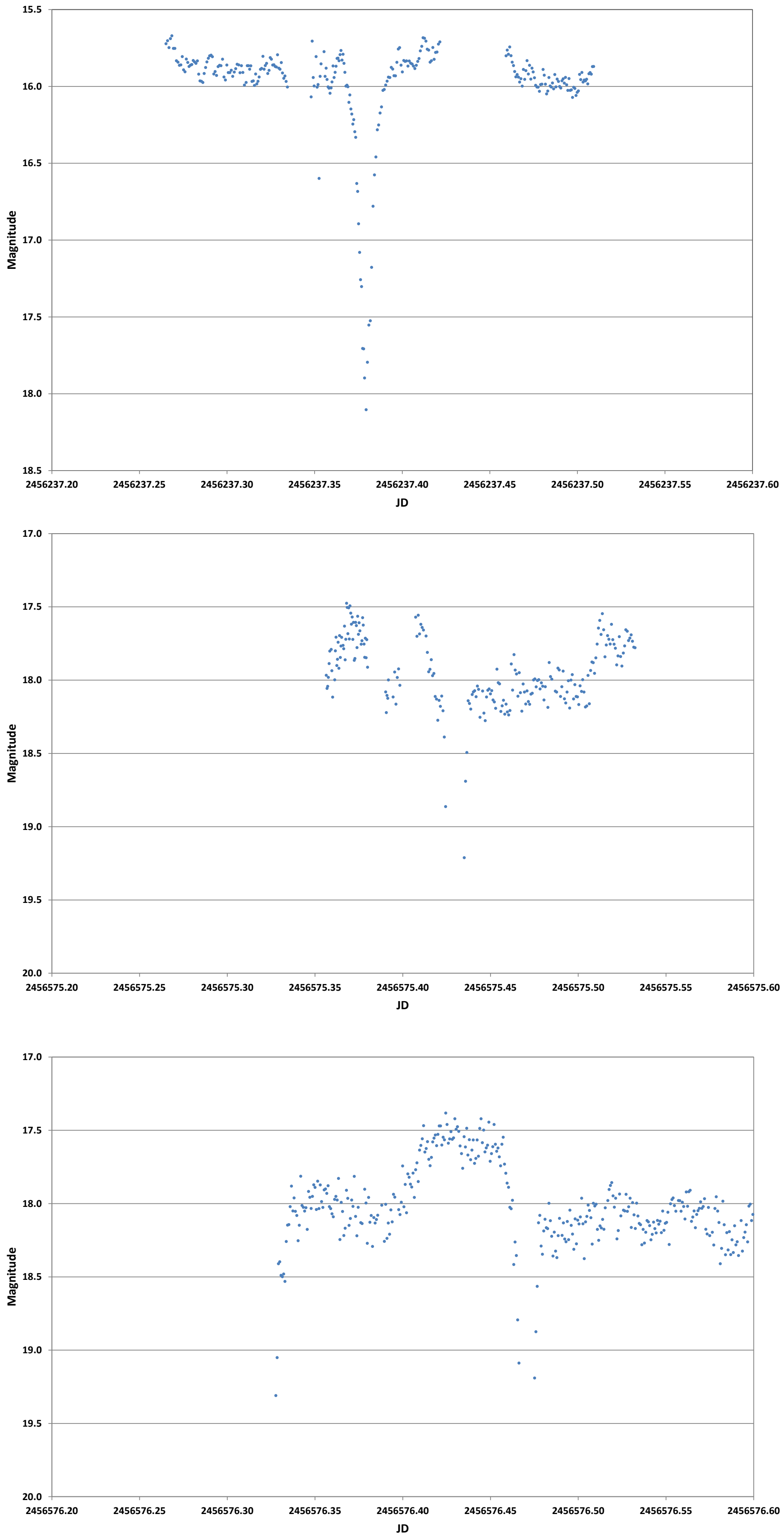

Figure 5: Examples of C-band photometry with small telescopes before and during the 2013/4 fade